\documentclass[sigconf,nonacm]{acmart}

\usepackage{booktabs}
\usepackage{graphicx}
\usepackage{amsmath}
\usepackage{enumitem}
\usepackage{rotating}
\usepackage{fancyhdr}

\renewcommand\footnotetextcopyrightpermission[1]{}

\fancypagestyle{plain}{%
  \fancyhf{}%
  \fancyhead{}%
  \fancyfoot[C]{\thepage}%
}
\fancypagestyle{firstpagestyle}{%
  \fancyhf{}%
  \fancyhead{}%
  \fancyfoot[C]{\thepage}%
}
\fancypagestyle{standardpagestyle}{%
  \fancyhf{}%
  \fancyhead{}%
  \fancyfoot[C]{\thepage}%
}
\begin{document}

\title{Visual RAG Toolkit: Scaling Multi-Vector Visual Retrieval\\with Training-Free Pooling and Multi-Stage Search}

\author{Ara Yeroyan}
\orcid{0009-0009-3605-9702}
\affiliation{%
  \institution{Independent Researcher}
  \country{Armenia}
}
\email{ar23yeroyan@gmail.com}

\begin{abstract}
Multi-vector visual retrievers (e.g., ColPali-style late interaction models) deliver strong accuracy, but scale poorly because each page yields thousands of vectors, making indexing and search increasingly expensive.
We present Visual RAG Toolkit, a practical system for scaling visual multi-vector retrieval with training-free, model-aware pooling and multi-stage retrieval.
Motivated by Matryoshka Embeddings, our method performs static spatial pooling--including a lightweight sliding-window averaging variant--over patch embeddings to produce compact tile-level and global representations for fast candidate generation, followed by exact MaxSim reranking using full multi-vector embeddings.

Our design yields a quadratic reduction in vector-to-vector comparisons by reducing stored vectors per page from thousands to dozens, notably without requiring post-training, adapters, or distillation.
Across experiments with interaction-style models such as ColPali and ColSmol-500M, we observe that over the limited (size) ViDoRe~v2 benchmark corpus 2-stage retrieval typically preserves NDCG and Recall @5/10 with minimal degradation, while substantially improving throughput (${\sim}4{\times}$ QPS); with sensitivity mainly at very large $k$.
The toolkit additionally provides robust preprocessing--high resolution PDF to image conversion, optional margin/empty-region cropping and token hygiene (indexing only visual tokens)--and a reproducible evaluation pipeline, enabling rapid exploration of two-, three-, and cascaded retrieval variants.
By emphasizing efficiency at common cutoffs (e.g., $k\!\le\!10$), the toolkit lowers hardware barriers and makes state-of-the-art visual retrieval more accessible in practice.
\end{abstract}

\maketitle
\thispagestyle{firstpagestyle}
\pagestyle{plain}

\section{Introduction}
\label{sec:intro}

Document retrieval has traditionally relied on text extraction pipelines--OCR, layout analysis, and text-based indexing--that discard the rich visual structure of real-world documents: charts, tables, infographics, coloured headings, and complex layouts.
Recent advances in vision-language models (VLMs) have opened an alternative: \emph{visual document retrieval}, where pages are embedded directly from their rendered images, bypassing OCR entirely~\cite{faysse2025colpali}.

The ColPali family of models~\cite{faysse2025colpali} adapts the late-interaction paradigm pioneered by ColBERT~\cite{khattab2020colbert} to the visual domain.
Each document page is encoded into a \emph{set} of patch embeddings by a VLM backbone (PaliGemma-3B for ColPali-v1.3, Qwen2-VL for ColQwen2.5~\cite{colqwen2024}, SmolVLM for ColSmol-500M~\cite{colsmol2024}), and query--document relevance is scored via MaxSim aggregation over all query--patch pairs.
This approach has achieved state-of-the-art results on the ViDoRe benchmark~\cite{mace2025vidorev2}, significantly outperforming text-only pipelines on visually rich documents.

\paragraph{\textbf{The scaling bottleneck.}}
Late-interaction accuracy comes at a steep computational cost.
A single page from \textbf{ColPali~(v1.3)} produces $D\!=\!1024$ vectors of dimension $d\!=\!128$; \textbf{ColQwen2.5-v0.2} accepts dynamic resolutions and produces up to ${\sim}768$ visual tokens per page (we observe $D\!\approx\!700{-}750$ in practice).

A typical query contains 8--12 words--for example, \emph{``What is the ESG risk assessment methodology?''} (8~words, $Q\!\approx\!10$ tokens\footnote{BPE tokenizers average ${\sim}1.3$ tokens/word for English (OpenAI guideline: 100 tokens $\approx$ 75 words).}).
Scoring one query--document pair requires $Q \times D$ inner products, each of dimension $d$; scanning $N$ pages multiplies this:
\begin{equation}
  \text{Cost}_\text{search} = Q \times D \times N \times d \quad\text{(multiply-adds)}
  \label{eq:cost}
\end{equation}

\noindent\textbf{Example.} For ColPali with $N\!=\!10{,}000$, $D\!=\!1024$, $Q\!=\!10$, $d\!=\!128$:
\[
  \underbrace{10 \times 1024 \times 10{,}000 \times 128}_{= 1.31 \times 10^{10}}\;\text{multiply-adds per query.}
\]
Reducing $D$ from $1024$ to $32$ pooled vectors (our row-mean pooling):
\[
  \underbrace{10 \times 32 \times 10{,}000 \times 128}_{= 4.10 \times 10^{8}}\;\text{multiply-adds per query}
  \longrightarrow  \mathbf{32{\times}\;\textit{reduction.}}
\]
The $d$ factor cancels in the ratio, so the saving \textbf{grows quadratically} with the compression ratio $D/D'$, regardless of dimension.

\smallskip
\noindent\textbf{Beyond search: index construction.}
The above concerns search alone.
For systems using \emph{HNSW indexing}, building the graph (with typical $M\!=\!16$, $\mathit{ef}_\text{c}\!=\!128$) requires $O\bigl(N \cdot \mathit{ef}_\text{c} \cdot \log N \cdot M\bigr)$ pairwise comparisons, where each comparison between two multi-vector points costs $O(D^2 \cdot d)$.
For $N\!=\!10{,}000$ pages with $D\!=\!1024$ and $d\!=\!128$, this amounts to \emph{trillions} of floating-point operations.
Reducing $D$ alleviates \emph{both} search and index construction costs simultaneously.

\smallskip
\noindent\textbf{Our approach vs.\ existing work.}
Existing efficiency techniques focus on \emph{quantization}~\cite{qdrant2024} (e.g., binary vectors) or engine-level optimisations like PLAID's centroid pruning, which reduce the \emph{cost per comparison} but not the \emph{number of comparisons}.
We target an orthogonal axis: \textbf{reducing the number of stored vectors per page} through training-free spatial pooling, then leveraging the compact representations for fast multi-stage retrieval.

\section{Methodology}
\label{sec:method}

We present \textbf{Visual RAG Toolkit}, an \textbf{end-to-end open-source system} for visual document retrieval that covers the full pipeline from PDF ingestion through indexed multi-stage search.
The toolkit is designed so that researchers, students, and practitioners can build and evaluate a complete visual RAG system on consumer hardware (see~\S\ref{sec:experiments}).
It provides three core contributions:

\begin{enumerate}
\item \textbf{Training-free, model-aware spatial pooling.}
Inspired by the idea behind Matryoshka Representation Learning~\cite{kusupati2022matryoshka}--that useful representations can be extracted at multiple granularities from a single embedding without additional training--we compress the full set of patch embeddings into compact multi-vector summaries (e.g., from ${\sim}1024$ to ${\sim}32$ vectors per page) using static spatial operations tailored to each model's architecture.

\item \textbf{Multi-stage retrieval} over Qdrant~\cite{qdrant2024} named vectors, using the compact pooled representations for fast candidate generation and the full patch embeddings for exact MaxSim reranking--all executed server-side in a single API call.

\item \textbf{Robust preprocessing and reproducible evaluation}: PDF-to-image conversion, optional empty-region cropping, token hygiene, and benchmark scripts that enable systematic ablation across models, pooling strategies, and retrieval configurations.
\end{enumerate}

\subsection{Token Hygiene}
\label{sec:token-hygiene}
VLMs emit several categories of non-visual tokens alongside the \emph{visual patch tokens} (the embeddings that each correspond to a spatial region of the input image).
These include:
(i)~\textbf{special tokens} such as CLS, BOS/EOS;
(ii)~\textbf{prompt/instruction tokens}, e.g., ``\emph{$\langle$bos$\rangle$Describe the image}'' prepended by ColPali~(v1.3);
and (iii)~\textbf{padding tokens} introduced during batch processing, where images of different sizes are padded to a uniform sequence length within a batch, producing trailing zero vectors.

Standard MaxSim treats all tokens equally, allowing non-visual tokens to act as \textbf{spurious high-similarity attractors} that inflate scores.
While filtering seems straightforward, it is notably \textbf{not done} by raw ViDoRe benchmark\footnote{ViDoRe~\cite{mace2025vidorev2} is a page-level visual document retrieval benchmark; see \S\ref{sec:data}.} submissions--models are evaluated with all tokens, including padding.

We \textbf{detect and strip non-visual tokens at index time}.
In practice, ColPali~(v1.3) retains $1024$ of $1030$ total tokens; ColQwen2.5 retains $720{-}768$ (mean $743$).
This reduces inner products (Eq.~\ref{eq:cost}) and improves quality--our ``clean'' 1-stage baseline sometimes exceeds published ViDoRe~v2 leaderboard scores, because non-visual tokens no longer distort MaxSim.
Cleaner vectors also make downstream pooling more reliable.

\subsection{Empty-Region Cropping}
\label{sec:cropping}
Document pages frequently contain blank margins, headers, and page numbers.
We optionally detect and remove low-variance border regions using row/column standard-deviation thresholds, with configurable page-number strip removal.
The tighter crop focuses encoder capacity on content, benefiting models with fixed input resolution (ColPali).
For dynamic-resolution models (ColSmol, ColQwen2.5), the benefit is twofold: not only does the encoder see more informative pixels, but a smaller cropped image also produces \emph{fewer patches and tiles}, directly reducing the number of stored vectors per page and the inner-product count at search time.

\subsection{Spatial Pooling Strategies}
\label{sec:pooling}
Our core insight is that the spatial structure of patch embeddings can be exploited to produce compact summaries \emph{without any training}.
We developed model-specific strategies iteratively, driven by the observation that different backbones require different approaches.
We describe them in the order we developed them.

\subsubsection{ColSmol-500M}
\label{sec:pool-colsmol}

\paragraph{\textbf{Tile-level mean pooling.}}
ColSmol's processor resizes each page to $512{\times}512$ pixels, partitions it into an $n_\text{rows}{\times}n_\text{cols}$ grid of tiles (each producing $P\!=\!64$ patch tokens), and appends one \emph{global tile}--a squeezed version of the entire original image--yielding $n_\text{rows} \cdot n_\text{cols} + 1$ tile groups (typically $12 + 1 = 13$) and ${\sim}832$ total patch tokens.
We mean-pool within each tile group to obtain one vector per tile:
\begin{equation}
  \mathbf{t}_i = \frac{1}{P}\sum_{p=1}^{P} \mathbf{x}_{(i,p)} \in \mathbb{R}^d, \quad i = 1,\ldots,n_\text{tiles}
  \label{eq:tile-pool}
\end{equation}
\noindent Result: ${\sim}832 \to {\sim}13$ vectors--a \textbf{64$\boldsymbol{\times}$ compression}.

\subsubsection{ColPali (v1.3)}
\label{sec:pool-colpali}

\paragraph{\textbf{Row-wise mean pooling.}}
ColPali uses a fixed $32{\times}32$ patch grid ($1024$ visual tokens, $d\!=\!128$).
We reshape tokens to a 2D grid and mean-pool across columns:
\begin{equation}
  \mathbf{r}_h = \frac{1}{W}\sum_{w=1}^{W} \text{grid}[h,w] \in \mathbb{R}^d, \quad h = 1,\ldots,H
  \label{eq:row-pool}
\end{equation}
\noindent Result: $1024 \to 32$ row vectors--a \textbf{32$\boldsymbol{\times}$ reduction}.

\paragraph{\textbf{Conv1d experimental pooling.}}
On top of these row vectors, we apply a \emph{uniform sliding window} of size $k\!=\!3$ with boundary extension, producing $N\!+\!2$ output vectors from $N$ input rows:
\begin{equation}
  y_i = \frac{1}{|W_i|}\sum_{j \in W_i} \mathbf{r}_j, \quad W_i = \bigl\{j : |j-(i\!-\!r)| \le r,\; 0 \le j < N\bigr\}
  \label{eq:conv1d}
\end{equation}
This adds inter-row context at negligible cost.
For ColPali, where \textbf{no learned local mixing} exists in the backbone, uniform averaging works well.

\subsubsection{ColQwen2.5 (v0.2)}
\label{sec:pool-colqwen}
ColQwen2.5~\cite{colqwen2024} builds on Qwen2-VL, which accepts images at variable aspect ratios and applies a learned \texttt{PatchMerger}: each $2{\times}2$ block of patch tokens is fused via LayerNorm $\to$ concatenation $\to$ MLP, reducing the grid from $H{\times}W$ to $H_\text{eff}{\times}W_\text{eff}$ ($H_\text{eff}\!\approx\!\lceil H/2\rceil$).
Because PatchMerger is a \textbf{learned spatial mixing} (not a simple average), each output token already encodes its $2{\times}2$ neighbourhood.

\smallskip
\noindent\textbf{Why ColPali's conv1d failed here.}
We initially applied the same conv1d approach that worked for ColPali (\S\ref{sec:pool-colpali}).
It \emph{degraded} ColQwen2.5's retrieval quality.
The cause: uniform averaging over already-mixed representations \textbf{double-smooths} spatial information, washing out discriminative features; the $N\!+\!2$ border extension further introduces artifacts where the backbone does not expect them.
This failure motivated a distinct pooling strategy.

\paragraph{\textbf{Weighted same-length smoothing (Gaussian / Triangular).}}
Instead of conv1d, we designed \textbf{same-length smoothing} ($N\!\to\!N$) with non-uniform weights:
\begin{equation}
  y_i = \frac{1}{Z_i}\sum_{j=i-r}^{i+r} w_{|j-i|}\,\mathbf{r}_j,
  \;\; Z_i = \!\!\!\sum_{\substack{j=i-r \\ 0 \le j < N}}^{i+r}\!\!\! w_{|j-i|}
  \label{eq:smooth}
\end{equation}
Boundary indices outside $[0,N)$ are skipped and weights re-normalised.
With $k\!=\!3$ ($r\!=\!1$):

\smallskip
\begin{itemize}[nosep,leftmargin=*]
\item \textbf{Gaussian}: $w_\delta = \exp(-\delta^2/2\sigma^2)$, $\sigma = \max(0.5,\,r/2)$; weights $\approx [0.61,\; 1.0,\; 0.61]$.
\item \textbf{Triangular}: $w_\delta = (r\!+\!1) - \delta$; weights $= [1,\; 2,\; 1]$.
\end{itemize}
\smallskip

\noindent Since PatchMerger already provides learned local context, only \textbf{gentle} smoothing is needed; Gaussian ($\sigma\!\approx\!0.5$) works best as its rapid decay preserves center-row identity.

\paragraph{\textbf{Adaptive row-mean pooling for dynamic resolution.}}
Since ColQwen2.5's grid $H_\text{eff}{\times}W_\text{eff}$ varies per image, we mean over columns (preserving vertical/reading-order structure), then adaptively downsample rows to at most $T$ vectors (default $T\!=\!32$) using evenly-spaced bins.
Pages with $H_\text{eff}\!<\!T$ are \emph{not} upsampled.

\subsection{Multi-Stage Retrieval}
\label{sec:retrieval}
Multi-stage retrieve-then-rerank pipelines are a well-established pattern in information retrieval: a cheap first stage (e.g., BM25 or a bi-encoder) retrieves a broad candidate set, and a more expensive model reranks it~\cite{khattab2020colbert}.
We apply the same principle \emph{within} the multi-vector paradigm: the cheap stage uses our compact pooled vectors, and the expensive stage uses the full patch embeddings--both stored in the same Qdrant collection as named vectors.

Concretely, each page is stored with:
\texttt{initial} (full multi-vector, ${\sim}700{-}1024$ vectors), \texttt{mean\_pooling} (row/tile-pooled, ${\sim}13{-}32$), experimental smoothed variants, and \texttt{global\_pooling} (single vector).
\textbf{2-stage retrieval} prefetches top-$K$ candidates via MaxSim on a compact named vector, then reranks with exact MaxSim on \texttt{initial}--executed entirely server-side via Qdrant's \texttt{prefetch}+\texttt{query} API, minimising round-trips.
A \textbf{3-stage} cascade adds a global-pooling prefetch before the pooled-vector stage.
The toolkit exposes all hyperparameters (prefetch-$K$, stage-1 vector choice, top-$k$, cascade depth) for systematic exploration of the accuracy--latency trade-off.

\section{Data and Evaluation Protocol}
\label{sec:data}

We evaluate on ViDoRe~v2~\cite{mace2025vidorev2}, a page-level visual document retrieval benchmark that emphasises realistic, non-extractive queries and diverse document types (charts, tables, infographics, multilingual content).
ViDoRe~v2 provides four English-language datasets.
We select three that are topically distinct:
\textbf{ESG Reports} (1538 multilingual pages, 227~queries),
\textbf{Biomedical Lectures} (1016 pages, 639~queries),
and \textbf{Economics Reports} (452 pages, 231~queries)--3006 pages total.
The 4th dataset covers the same ESG domain in English only; we exclude it to avoid topical redundancy and keep the distractor analysis clean.

\paragraph{\textbf{Evaluation scopes and the distractor experiment.}}
We evaluate each configuration in two scopes:

(i)~\textbf{Per-dataset}: each query searches only its own corpus (comparable to the official ViDoRe leaderboard).
This is the natural starting point--it establishes that 2-stage retrieval preserves accuracy on each domain independently.

(ii)~\textbf{Union (distractor)}: all 3006 pages are merged into a single Qdrant collection, so each query must find its relevant pages among cross-dataset distractors.
Since 1-stage cost scales linearly with $N$ (Eq.~\ref{eq:cost}) while 2-stage reranking is capped at $K$ candidates, this scope directly tests whether the speedup advantage grows with corpus size.

\paragraph{\textbf{Metrics.}}
NDCG and Recall @$k \in \{5, 10, 100\}$; throughput (QPS).

\section{Experiments}
\label{sec:experiments}

The toolkit exposes dozens of configurable parameters: number of retrieval stages, prefetch-$K$, stage-1 vector choice, pooling kernel, cropping thresholds, and more.
We explore representative configurations that isolate the effect of our pooling and multi-stage retrieval, keeping other variables fixed.

\paragraph{\textbf{Model and hardware selection.}}
We deliberately select models that are practical on consumer hardware:
ColSmol-500M~\cite{colsmol2024} (500M parameters, tile grid, ${\sim}832$ patches),
\textbf{ColPali-v1.3}~\cite{faysse2025colpali} (3B, fixed $32{\times}32$ grid, 1024 patches),
and \textbf{ColQwen2.5-v0.2}~\cite{colqwen2024} (3B, dynamic resolution, PatchMerger).
All vectors are stored in FP16; Qdrant collections are fully in-RAM with no HNSW index.
The 500M--3B parameter range is chosen intentionally: these models run comfortably on a single consumer GPU or Apple Silicon Mac, making state-of-the-art late-interaction visual retrieval accessible to researchers, students, and practitioners without datacenter resources.
Combined with our pooling and multi-stage retrieval, a complete visual RAG system--from PDF ingestion through indexed search--can be built and evaluated on a laptop.

\paragraph{\textbf{Baselines.}}
Two baselines are relevant.
The first is the \emph{official ViDoRe v2 leaderboard} score for each model (Table~\ref{tab:vidore}), which uses the raw model output (all tokens, including padding and special tokens) without any preprocessing.
The second--and our primary comparison--is \emph{1-stage full}: exact MaxSim over all stored patch embeddings in our collection, \emph{after} applying our token hygiene and optional cropping (\S\ref{sec:token-hygiene}--\ref{sec:cropping}).
Notably, even though we store vectors in FP16 (lower precision than the leaderboard's FP32), our 1-stage baseline occasionally \emph{exceeds} the official scores (compare Tables~\ref{tab:vidore} and~\ref{tab:main})--demonstrating that token hygiene and cropping can matter more than numerical precision.
We report all 2-stage results (prefetch $K\!=\!256$, top-$100$) relative to this stronger baseline.

\section{Results}
\label{sec:results}

\begin{table}[t]
\centering
\caption{\footnotesize Official ViDoRe v2 leaderboard (per-dataset, no preprocessing).}
\label{tab:vidore}
\footnotesize
\setlength{\tabcolsep}{2pt}
\begin{tabular}{@{}ll ccccc@{}}
\toprule
& & \textbf{N@5} & \textbf{N@10} & \textbf{R@5} & \textbf{R@10} & \textbf{R@100} \\
\midrule
\emph{ColPali} & ESG  & .549 & .574 & .539 & .643 & .921 \\
\emph{v1.3}    & Bio  & .546 & .588 & .575 & .702 & \textbf{.914} \\
               & Econ & .486 & .484 & .272 & .400 & .878 \\
\midrule
\emph{ColQwen} & ESG  & \textbf{.664} & \textbf{.682} & \textbf{.718} & \textbf{.776} & \textbf{.925} \\
\emph{2.5}     & Bio  & \textbf{.592} & \textbf{.631} & \textbf{.612} & \textbf{.729} & .901 \\
               & Econ & \textbf{.533} & \textbf{.528} & \textbf{.297} & \textbf{.420} & \textbf{.904} \\
\bottomrule
\end{tabular}
\end{table}

\begin{table}[t]
\centering
\caption{\footnotesize Our results (union scope, 3006 pages, with token hygiene).}
\label{tab:main}
\footnotesize
\setlength{\tabcolsep}{1.8pt}
\begin{tabular}{@{}r@{\;\;}l l ccccc c@{}}
\toprule
& & & \textbf{N@5} & \textbf{N@10} & \textbf{R@5} & \textbf{R@10} & \textbf{R@100} & \textbf{QPS} \\
\midrule[0.08em]
\smash{\raisebox{-1.2em}{\rotatebox[origin=c]{90}{\scriptsize\textbf{ColPali v1.3}\quad}}}
& \emph{1stage base} & ESG  & .551 & .573 & .570 & .661 & .908 & 0.28 \\
& & Bio  & .566 & .601 & \textbf{.611} & .713 & \textbf{.911} & 0.27 \\
& & Econ & .488 & .490 & .244 & .401 & .855 & 0.31 \\[2pt]
\cmidrule(l){2-9}
& \emph{2stage Conv1d} & ESG  & \textbf{.559}{\tiny+.01} & \textbf{.576}{\tiny .00} & \textbf{.576}{\tiny+.01} & .662{\tiny .00} & .818{\tiny$-$.09} & 1.27 \\
& & Bio  & .566{\tiny .00} & .601{\tiny .00} & .610{\tiny .00} & .712{\tiny .00} & .902{\tiny$-$.01} & 1.37 \\
& & Econ & .486{\tiny .00} & .489{\tiny .00} & .243{\tiny .00} & .401{\tiny .00} & .834{\tiny$-$.02} & 1.39 \\ \\
\midrule[0.08em]
\smash{\raisebox{-1.2em}{\rotatebox[origin=c]{90}{\scriptsize\textbf{ColQwen 2.5}\quad}}}
& \emph{1stage base} & ESG  & .509 & .564 & .538 & \textbf{.704} & .856 & 0.31 \\
& & Bio  & \textbf{.575} & \textbf{.612} & .605 & \textbf{.718} & .880 & 0.32 \\
& & Econ & \textbf{.572} & .551 & .308 & .432 & \textbf{.873} & 0.33 \\[2pt]
\cmidrule(l){2-9}
& \emph{2stage Gauss.} & ESG  & .513{\tiny .00} & .560{\tiny .00} & .549{\tiny+.01} & .690{\tiny$-$.01} & .788{\tiny$-$.07} & 1.15 \\
& & Bio  & .574{\tiny .00} & .611{\tiny .00} & .605{\tiny .00} & .716{\tiny .00} & .865{\tiny$-$.02} & 1.28 \\
& & Econ & \textbf{.572}{\tiny .00} & \textbf{.552}{\tiny .00} & \textbf{.309}{\tiny .00} & \textbf{.433}{\tiny .00} & .821{\tiny$-$.05} & 1.25 \\ \\
\midrule[0.08em]
& \emph{1stage base} & ESG  & .404 & .457 & .460 & .620 & \textbf{.961} & 0.50 \\
& & Bio  & .390 & .420 & .431 & .519 & .778 & 0.50 \\
& & Econ & .323 & .327 & .165 & .268 & .692 & 0.52 \\[2pt]
\cmidrule(l){2-9}
\smash{\raisebox{-0.6em}{\rotatebox[origin=c]{90}{\scriptsize\textbf{ColSmol 500M}\quad}}}
& \emph{2stage tiles} & ESG  & .369{\tiny$-$.04} & .401{\tiny$-$.06} & .405{\tiny$-$.05} & .503{\tiny$-$.12} & .658{\tiny$-$.30} & 1.32 \\
& & Bio  & .381{\tiny$-$.01} & .406{\tiny$-$.01} & .413{\tiny$-$.02} & .490{\tiny$-$.03} & .677{\tiny$-$.10} & 1.50 \\
& & Econ & .329{\tiny+.01} & .325{\tiny .00} & .171{\tiny+.01} & .261{\tiny$-$.01} & .636{\tiny$-$.06} & 1.54 \\[2pt]
\cmidrule(l){2-9}
& \emph{3stage casc.} & ESG  & .371{\tiny$-$.03} & .408{\tiny$-$.05} & .406{\tiny$-$.05} & .520{\tiny$-$.10} & .678{\tiny$-$.28} & 0.86 \\
& & Bio  & .371{\tiny$-$.02} & .397{\tiny$-$.02} & .399{\tiny$-$.03} & .477{\tiny$-$.04} & .655{\tiny$-$.12} & 0.88 \\
& & Econ & .332{\tiny+.01} & .330{\tiny .00} & .173{\tiny+.01} & .266{\tiny .00} & .688{\tiny .00} & 0.88 \\
\bottomrule
\end{tabular}
\end{table}

Table~\ref{tab:main} reports our union-scope results (N = NDCG, R = Recall in the table headers).
We conducted many more experiments than shown--varying pooling kernels, prefetch-$K$, cascade depth, and cropping--and report a representative subset per model.

\paragraph{\textbf{ColPali and ColQwen2.5: ${\sim}4{\times}$ faster, nearly lossless.}}
For both 3B models, 2-stage retrieval achieves \textbf{3.8--4.5$\boldsymbol{\times}$ QPS} while preserving N@5, N@10, R@5, and R@10 within ${\pm}0.01$ of the 1-stage baseline.
Degradation appears only at R@100 ($-$0.02 to $-$0.09), where the 256-candidate prefetch window limits coverage--acceptable in practice, since RAG applications typically use $k\!\le\!10$.

\paragraph{\textbf{ColSmol-500M: small models degrade more.}}
ColSmol shows larger drops (up to $-$0.30 R@100), suggesting that sub-1B models lack sufficient representational capacity for pooling to remain lossless.
The 3-stage cascade recovers some recall but at lower QPS.

\paragraph{\textbf{Throughput.}}
In \emph{per-dataset} evaluation (each dataset searched independently, 452--1538 pages), 2-stage yields ${\sim}2{\times}$ QPS.
In the \emph{union} setting (all 3006 pages combined as distractors), speedup grows to ${\sim}4{\times}$.
This is consistent with the quadratic cost reduction from \S\ref{sec:intro} (Eq.~\ref{eq:cost}): as $N$ grows, 1-stage cost increases linearly while 2-stage reranking is capped at $K\!=\!256$ candidates.
Our corpus is modest; the $2{\times}{\to}4{\times}$ trend with just a $3{\times}$ increase in $N$ suggests that larger collections will see even greater gains.

\paragraph{\textbf{Pooling kernel selection.}}
For ColQwen2.5, conv1d degraded quality (\S\ref{sec:pool-colqwen}); Gaussian slightly outperformed Triangular.
For ColPali, conv1d and tile-based pooling performed comparably.

\section{Demo System}
\label{sec:demo}
A Streamlit demo\footnote{\url{https://huggingface.co/spaces/Yeroyan/visual-rag-toolkit}} lets users upload PDFs, index into a free Qdrant Cloud account using ColPali-family models, and query with visual retrieval (PDF$\to$images$\to$cropping$\to$pooling$\to$indexing$\to$multi-stage search$\to$results).
The open-source package\footnote{\url{https://github.com/Ara-Yeroyan/visual-rag-toolkit}} (\texttt{pip install visual-rag-toolkit}) provides a CLI, modular SDK, and benchmark scripts to reproduce all experiments.

\section{Conclusion}
\label{sec:conclusion}

We presented Visual RAG Toolkit, an end-to-end open-source system that makes multi-vector visual document retrieval practical on consumer hardware.
Our central contribution is \textbf{training-free, model-aware spatial pooling} that reduces stored vectors per page from ${\sim}1024$ to ${\sim}32$, yielding a \textbf{quadratic reduction} in inner-product computations (Eq.~\ref{eq:cost}).
Combined with multi-stage retrieval--cheap prefetch on pooled vectors, exact MaxSim rerank on full embeddings--the toolkit achieves up to ${\sim}4{\times}$ throughput improvement with negligible quality loss at practical cutoffs ($k\!\le\!10$).

The quality degradation we observe concentrates at Recall@100, a regime rarely needed in production RAG systems and chatbots where $k\!\le\!10$ is standard.
For the 3B models (ColPali-v1.3 and ColQwen2.5), retrieval metrics at $k\!\le\!10$ remain within ${\pm}0.01$ of the uncompressed baseline--a strong indication that spatial pooling preserves the information most relevant for practical retrieval.
The sub-1B model (ColSmol-500M) shows larger drops, pointing to a representational capacity threshold below which aggressive pooling becomes lossy.

Crucially, the speedup advantage \emph{grows with corpus size}: from ${\sim}2{\times}$ in per-dataset evaluation to ${\sim}4{\times}$ in our union (distractor) setting with just a $3{\times}$ increase in $N$.
This trend is consistent with the quadratic cost analysis of \S\ref{sec:intro} and suggests that larger real-world collections will see even greater efficiency gains.

\paragraph{\textbf{Limitations and future work.}}
Our pooling strategies are model-specific: each backbone's tokenisation and spatial processing requires a tailored approach.
Specifically, fixed-grid models (ColPali) use conv1d sliding-window pooling, PatchMerger models (ColQwen) require weighted same-length smoothing (Gaussian/Triangular), and tile-based models (ColSmol) use tile-level mean pooling.
However, most current visual retrieval models share one of these three architectural patterns, so the existing strategies cover the majority of the ecosystem.
For entirely new architectures, the toolkit's modular design makes it straightforward to implement a new pooling function without changing the retrieval or evaluation pipeline.

Additional directions for future work include:
(i)~improving pooling quality to the point where 1-stage retrieval on pooled embeddings alone becomes viable--achieving not only speed but also significant storage reduction by eliminating the need to store full patch vectors;
(ii)~exploring learned pooling (e.g., lightweight adapters) that could close the quality gap for small models;
and (iii)~combining our vector-count reduction with orthogonal techniques such as quantization and HNSW pruning for multiplicative efficiency gains.

\bibliographystyle{ACM-Reference-Format}
\bibliography{references}

\end{document}